\documentclass{aastex631}
\usepackage{url}
\usepackage{bm}
\usepackage{tabularx}
\usepackage[none]{hyphenat}

\newcommand\clearrow{\global\let\rowmac\relax}
\usepackage{float}
\usepackage{graphicx}
\usepackage{verbatim} 
\usepackage{xcolor,soul}
\definecolor{pastelyellow}{rgb}{0.99, 0.99, 0.59}

\begin{document}

\title{Strong Photospheric Heating Indicated by Fe\,I 6173\,\AA~Line Emission During White-Light Solar Flares}

\author{Samuel Granovsky}
\affiliation{Department of Physics, New Jersey Institute of Technology, Newark, NJ 07102, USA}

\author{Alexander G. Kosovichev}
\affiliation{Department of Physics, New Jersey Institute of Technology, Newark, NJ 07102, USA}
\affiliation{NASA Ames Research Center, Moffett Field, Mountain View, CA 94040, USA}

\author{Viacheslav M. Sadykov}
\affiliation{Department of Physics \& Astronomy, Georgia State University, Atlanta, GA 30302}

\author{Graham S. Kerr}
\affiliation{Department of Physics, Catholic University of America, 620 Michigan Avenue, Northeast, Washington, DC 20064, USA}
\affiliation{NASA Goddard Space Flight Center, Heliophysics Science Division, Code 671, 8800 Greenbelt Road, Greenbelt, MD 20771, USA}

\author{Joel C. Allred}
\affiliation{NASA Goddard Space Flight Center, Heliophysics Science Division, Code 671, 8800 Greenbelt Road, Greenbelt, MD 20771, USA}


\begin{abstract}
Between 2017 and 2024, the Helioseismic and Magnetic Imager (HMI) onboard the Solar Dynamics Observatory has observed numerous white-light solar flares (WLFs). HMI spectro-polarimetric observations of certain WLFs, in particular the X9.3 flare of September 6, 2017, reveal one or more locations within the umbra or along the umbra/penumbra boundary of the flaring active region where the Fe\,I 6173\,\AA~line briefly goes into full emission, indicating significant heating of the photosphere and lower chromosphere. For five flares featuring Fe\,I 6173\,\AA~line-core emission, we perform spectro-polarimetric analysis using HMI 90\,s cadence Stokes data. For all investigated flares, line-core emission is observed to last for a single 90\,s frame and is either concurrent with or followed by an increase in the line continuum intensity lasting one to two frames (90\,s\,--\,180\,s). Additionally, permanent changes to the Stokes Q, U, and/or V profiles were observed, indicating long-lasting non-transient changes to the photospheric magnetic field. These emissions coincided with local maxima in hard X-ray emission observed by Konus-Wind, as well as local maxima in the time derivative of soft X-ray emission observed by GOES 16-18. Comparison of the Fe\,I 6173\,\AA~line profile synthesis for the ad-hoc heating of the initial empirical VAL-S umbra model and quiescent Sun (VAL-C-like) model indicates that the Fe\,I 6173\,\AA~line emission in the white-light flare kernels could be explained by the strong heating of initially cool photospheric regions.
 
\end{abstract}


\section{Introduction} \label{sec:Intro}

Solar flares are powerful, localized electromagnetic radiation bursts in the solar atmosphere, which release vast amounts of energy across the electromagnetic spectrum. While flares are mainly associated with high X-ray and (extreme-) UV emission, certain flares are accompanied by an increase in visible continuum intensity and are thus referred to as white-light flares (WLFs). While white-light enhancement relative to the initial photospheric luminosity is typically greatest during stronger flares, WLFs have been observed as low as GOES class C1.6 \citep{Hudson2006, Jess2008}.  

The origin of WLFs, and, in particular, the compact and short-lived WLF kernels (or cores), has been a subject of long-standing debates \citep[see,][and references therein]{Machado1986,Neidig1989,Fletcher2011,Kerr2014, Prochazka2019}. Early observations showed that the WLF cores are extremely small with a characteristic size of 1.5\,--\,3 arcsec (1\,--\,2\,Mm) when observed in central parts of the solar disk. However, more recent observations by Hinode have revealed even smaller WLF cores with sizes less than 1\,Mm \citep{Isobe2007}. Their area is less than 1\% of the flare area observed for the same flares in H$\alpha$. The WLF cores are observed in the penumbra areas in the vicinity of the magnetic polarity inversion line \citep{vSvestka1970}. 


Although observations from the Helioseismic and Magnetic Imager (HMI) on Solar Dynamics Observatory \citep{Scherrer2012} showed that enhancements of the continuum in the vicinity of the Fe\,I\,6173\,\AA~line are often observed during WLFs, emission in the line-core that should be considered as a separate phenomenon is rather rare. In the HMI data, such events can be traced in all observables, including the Doppler shift and line-of-sight magnetic field. However, due to rapid and strong variations of the line profile, the HMI 45-sec observing sequence and the inversion algorithm do not provide reliable estimates of magnetic field and plasma velocity in these events \citep{Sadykov2020}. 

Spectroscopic and photometric observations suggested that the mostly likely source of the observed white-light emission is the heating of the low chromosphere and upper photosphere to high temperatures \citep[e.g.][]{Machado1986,Neidig1989,Kerr2014}. However, explaining the WLF events in the framework of the standard thick-target flare model, which assumes the flare energy release in the form of precipitating electron beams, is difficult because the energy carried by electrons is predominantly deposited in the upper chromosphere. There is insufficient power carried by the highest energy electrons to produce meaningful effects in the photosphere.  Radiative hydrodynamics simulations (RHD) have suggested that the white-light emission can be produced by free-bound hydrogen emission in a thin 'chromospheric condensation' layer formed behind the downward propagating radiative shock \citep{Livshits1981,Kosovichev1986,Kowalski2015}, perhaps, with some contribution of radiatively heated photospheric layers \citep{Gan2000}. In addition RHD modeling of electron beam heating with various low-energy cut-off values, $E_c$, using the RADYN code \citep{1997ApJ...481..500C,Allred2005,2015ApJ...809..104A}, showed that the electron beams with $E_c \ge 100$\,keV can deposit their energy in the low chromosphere and explain Type II  WLFs\footnote{Type 1 WLFs are co-temporal with impulsive hard X-rays and have been associated with Balmer jumps and strong Balmer line response, whereas Type II WLFs are more gradual in nature and do not always have strong response of chromospheric lines.} \citep{Prochazka2019}. However, the electron beams with such high low energy cut-off are not observed.

It is important to note that the WLF events observed by the HMI instrument are often accompanied by substantial rapid variations in the Fe\,I\,6173\,\AA~line profile, which is a deep photospheric line, and excitation of helioseismic acoustic waves ('sunquakes') \citep{Sharykin2020}. The excitation of sunquakes provides unambiguous evidence that the flare hydrodynamic impacts affect the solar photosphere and, in some cases, even the subphotospheric layers on the Sun \citep{Stefan2020,Lindsey2020,Stefan2022}. The HMI observations of the X1.5 flare of May 10, 2022, which produced strong sunquakes, showed emission in the core of the Fe\,I\,6173\,\AA~line in the sunquake sources coinciding with the bright continuum emission cores \citep{Kosovichev2023}. The analysis of the Stokes profiles at the emission cores revealed impulsive and permanent changes in the vertical and horizontal magnetic field components, unambiguously confirming previous indications of magnetic field changes in solar flares \citep{Zvereva1970,Zirin1981,Wang1994}. Comparisons with the RADYN models showed that the line core emission can be explained by proton beams with the cut-off energy $E_c \ge 500$\,keV, but stronger sunquake-like helioseismic impacts are provided by proton beams with $E_c \le 100$ keV \citep{Sadykov2024RADYN}. The idea that the flare white-light emission is produced by deeply penetrating protons was suggested previously \citep[e.g.][]{vSvestka1970,Simnett1986}. However, a consistent model that would explain the X-ray and optical observations of the WLF cores and account for both accelerated electrons and protons, has not yet been produced.

In this paper, we present a spectro-polarimetric analysis of five sunquake-producing white-light flares that took place between solar cycles 24 and 25. These flares are characterized by instances of the Fe\,I 6173\,\AA~line briefly going into full emission, a phenomenon which can not be fully accounted for by the proton beam simulations studied by \cite{Kerr2023} and \cite{Sadykov2024RADYN}, both of which used a plage-like pre-flare atmosphere. In particular, we examine the SOL2017-09-06T11:53 X9.3 GOES-class flare which featured the greatest number and intensity of Fe\,I 6173\,\AA~line emission. For the investigated flare events, all but one observed instance of line emission occurred above sunspot umbra/penumbra. Considering the disparity between proton beam cutoff energies which encourage line-core emission and sunquakes for the aforementioned models, the observations potentially suggest that cooler umbral/penumbral temperature profiles may allow for stronger line-core emission at cutoff energies which generate strong sunquakes. To provide additional evidence for this, we perform radiative transfer modeling of the Fe\,I 6173\,\AA~line for our existing proton beam simulations results, but with ad-hoc modifications to the temperature profiles guided by the VAL-S umbra and VAL-R penumbra models \citep{Fontenla2006}. While not  self-consistent, this experiment reveals that photospheric temperature profiles resembling those above sunspot umbrae may lead to the Fe\,I 6173\,\AA~line entering full emission. Conversely, profiles that correspond to those found above hotter sunspot penumbrae and quiet sun regions may be more likely to result in partial line core emission which does not exceed continuum intensity. The experiments serve to provide additional support to the idea that lower photospheric temperatures may increase the likelihood of achieving line reversal due to a sunquake-producing proton beam. These experiments also demonstrate the need to perform self-consistent RADYN modeling of electron, proton, or multi-species beam-driven flares using realistic umbral/penumbral pre-flare stratification.

\newpage

\section{HMI Observations} \label{sec:Res}

The HMI instrument collects spectro-polarimetric data by capturing linear and circular polarized narrow-band images at six wavelengths across the Fe\,I 6173\,\AA~line using two cameras \citep{Couvidat2012HMI, Schou2012HMI}. The images, with a resolution of $4096 \times 4906$ pixels, cover the entire disk at a sampling rate of 0.5 arcsec per pixel giving a spatial resolution of $\sim$1 arcsec. Polarized images are acquired every 3.75\,s, with the complete spectro-polarimetric measurement sequence taking 90\,s for linear polarization and 45\,s for circular polarization. These images are sequenced to minimize systematic errors caused by line variations during observing cycles and are then combined into the Stokes line profiles. Consequently, HMI observations deliver the full Stokes line profiles for the Fe\,I 6173\,\AA~line with a spectral resolution of 68 m\AA~, a spatial resolution of $\sim$1 arcsec, and a temporal resolution of 90\,s. {The measured Stokes profiles are then used to determine the vector magnetic field and LOS velocity through a simplified Milne-Eddington inversion procedure. However, due to the rapid variations in the line profile during flares, this approach may not reliably capture flare-related velocity and magnetic field variations \citep{Svanda2018, Kosovichev2023}, and detailed modeling of the HMI observational procedure is required \citep{Sadykov2020}. Therefore, we report transient features observed during the investigated flares; however, we caution that the Stokes profile variations may be affected by the line-profile sampling sequence of the HMI instrument. The line profile is recorded in filtergrams at six wavelengths across the line in four polarizations by two cameras, with the whole observing sequence taking 90\,s. The filtergrams are interpolated to the central time of the observing sequence and combined in the Stokes profiles. In addition, some of the observed long-term variations, particularly permanent changes in the magnetic field, may be influenced by the evolution of background features relative to a fixed coordinate system.} 

For each flare, 90\,s cadence HMI Stokes image data for 13 to 15 frames (representing a time span of 19.5 to 22.5 minutes) are analyzed while ensuring immediate pre-flare and post-flare conditions are included. Due to the 90\,s cadence, these emission events can not be effectively resolved temporally as particle beam heating lasts for significantly shorter than this. These data are retrieved from the Joint Science Operations Center (JSOC) database. After mapping to helioprojective coordinates, the images are cropped around the flaring region and are tracked relative to the first frame in the time-series based on the local differential rotation rate to allow for temporal analysis. The Stokes I images are then normalized to pre-flare continuum intensity, and Stokes Q, U, and V are normalized to pre-flare continuum intensity for a patch of quiet sun. Pre-flare continuum intensity is approximated using the mean of filtergram channel 1 and 6 data corresponding to $\lambda_0 \pm 170$m\AA~for the first image in the time-series. However, due to the low spectral sampling rate, the complex line shape within magnetized regions, and the fact that the full absorption line is not captured and is often off-center in HMI data, it is not possible to reliably estimate the true continuum intensity value \citep{Svanda2018}. To determine locations where the line goes into emission, line-core emission maps representing the line-core to line-wing (CtW) ratio are used. These maps are computed by dividing the mean of the two Stokes I maps corresponding to $\lambda_0 \pm 34$\,m\AA~(channels 3 and 4) by the mean of the two Stokes I maps corresponding to $\lambda_0 \pm 170$\,m\AA~(channels 1 and 6). Due to SDO's orbital motion (geosynchronous) and the Sun's rotation, the absolute wavelength bandpass for each filtergram channel varies over time resulting in an inconsistent CtW ratio definition relative to the true position of the line core. In addition, the complex line shapes which often occur within sunspot umbrae and flaring regions make it difficult to compute a robust estimate for the cropped portion of the line wing, meaning a reliable average intensity value of the line-wings at an equal distance from the line-core can not be found. Therefore, the CtW ratio as presented in this paper is not a metric which can be used to accurately compare the relative core emission strength between flares, and is instead used as an approximate method to find locations where the line goes into emission and only approximate differences in emission. However, for this particular set of observations, peaks in both line core absorption and emission exclusively occur in filtergram channels 3 and 4, so the CtW ratio metric always contains the line-core within its line-core window. In addition, the observation time-frame (roughly 20 minutes) is short enough to allow for relative comparisons of the line shape and position during the span of a single flare line-emission event.

Using this procedure, twenty-one flares observed by SDO between 2016-11-30 and 2024-03-28 with associated sunquakes from strong photospheric impacts were analyzed. Out of these twenty-one events, all featured instances of increased white-light emission. However, only five were found to have instances of the line going into either full emission or line-core emission comparable to or exceeding line wing intensity. The five events featuring significant line core emission are further discussed below. For each instance of significant line-core emission, the mean Stokes parameters of a 3x3 pixel area representing a 1.5 arcsec\textsuperscript{2} area centered on the location with a local maximum CtW ratio is used for analysis. 

\begin{figure*}[t]
	\centering
	\includegraphics[width=0.8\textwidth]{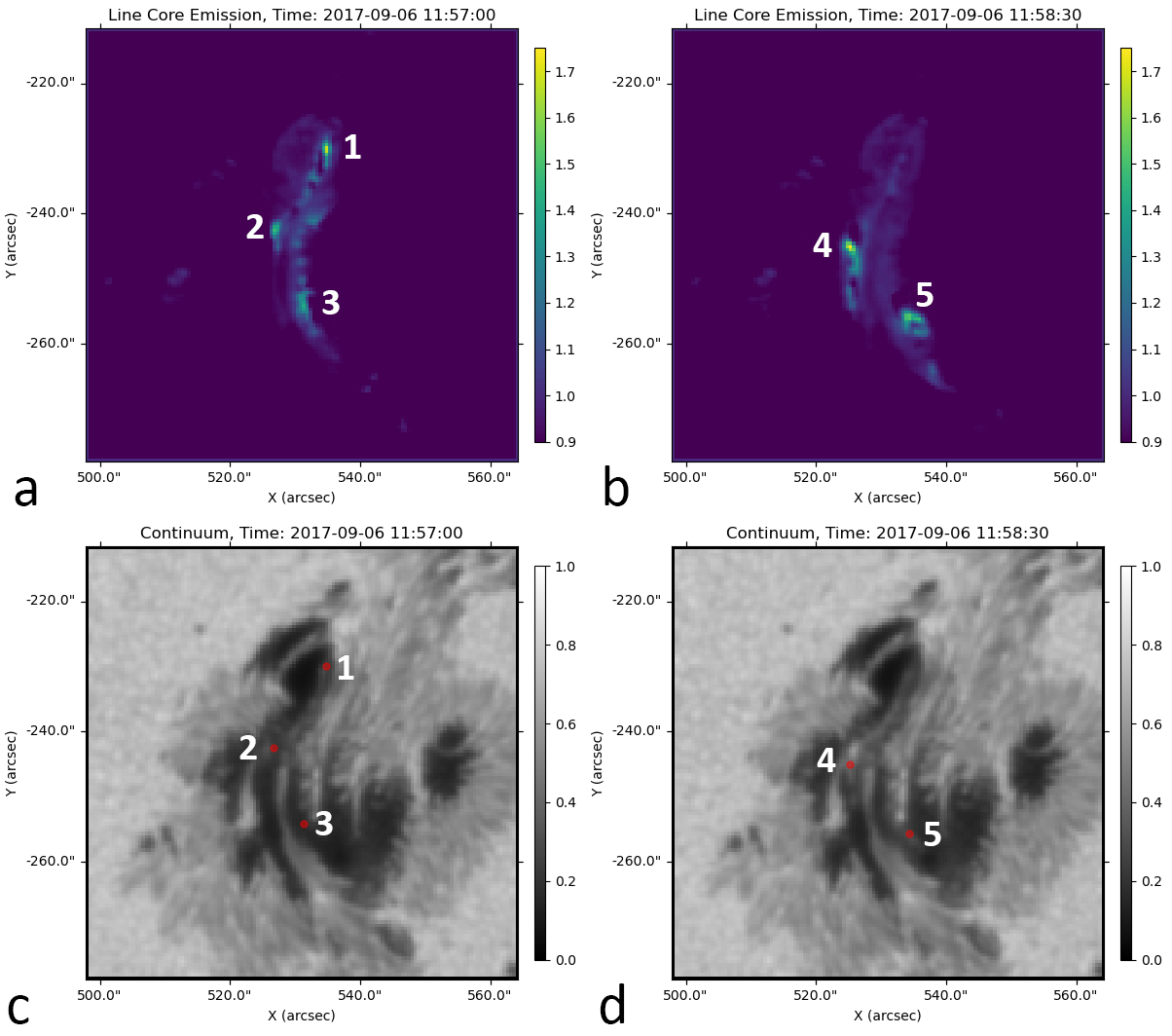}
	\caption{The Fe\,I 6173\,\AA~Stokes I line-core to line-wing (CtW) ratio observed by the HMI instrument during the 2017-09-06 X9.3 flare. Panels a and b show CtW ratio during maximum emission. Points 1\,--\,5 correspond to the five locations with strongest emission whose Stokes profiles are analyzed. Panels c and d show the same locations in red on the corresponding normalized HMI continuum maps.} \label{X93_CtW}
	\vspace{0.1cm}
\end{figure*}

\begin{figure*}[b]
	\centering	\includegraphics[width=0.8\textwidth]{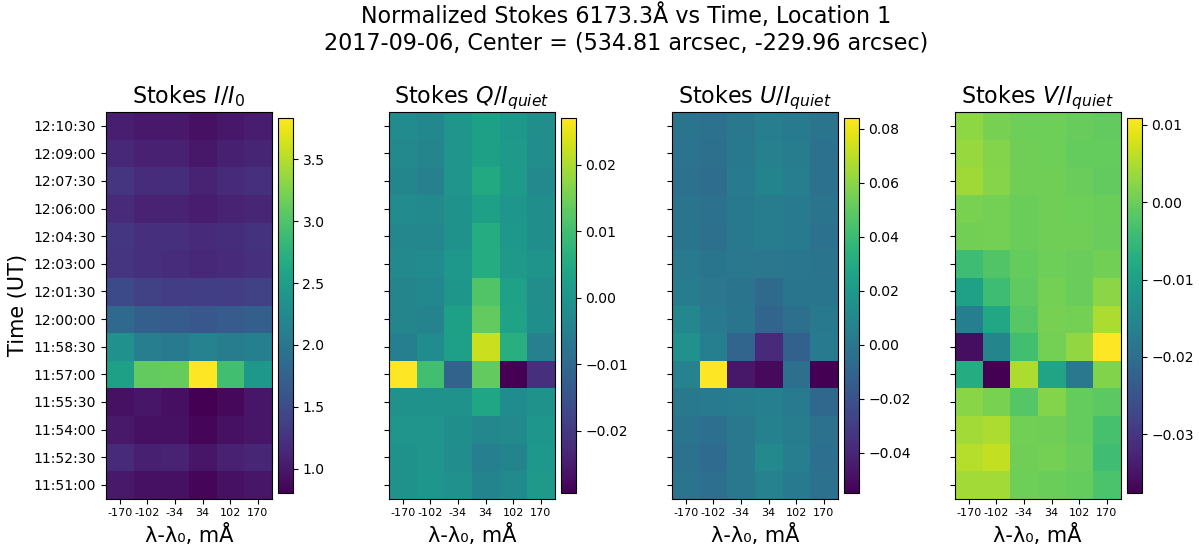}
	\caption{Fe\,I 6173\,\AA~Stokes parameters vs time for the 2017-09-06 X9.3 flare at location 1.}
    \label{X93_S1}
	\vspace{0.1cm}
\end{figure*}

\begin{figure*}
	\centering
	\includegraphics[width=0.8\textwidth]{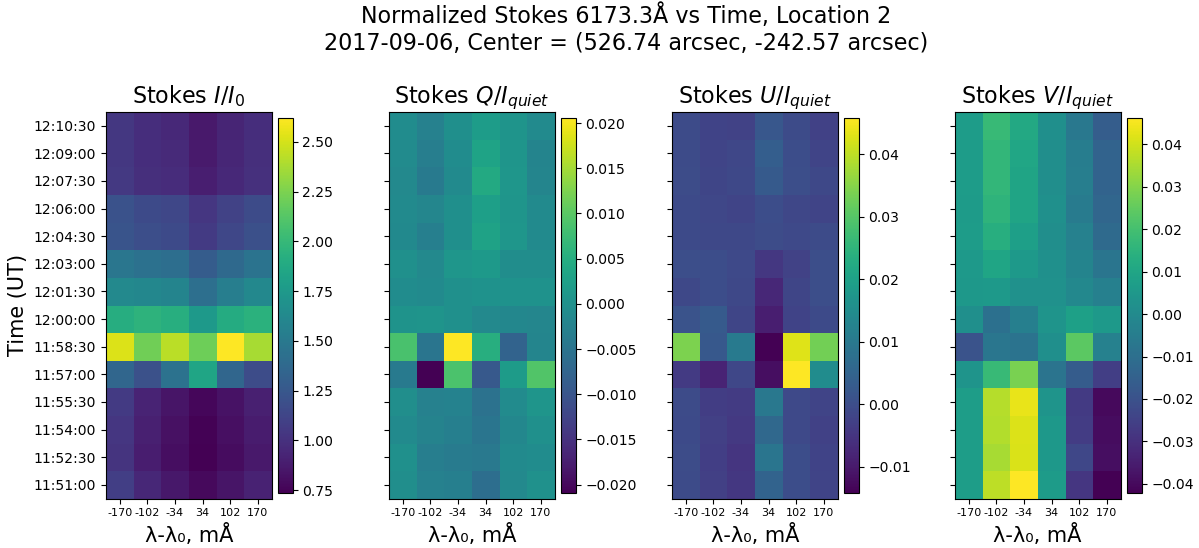}
	\caption{Fe\,I 6173\,\AA~Stokes parameters vs time for the 2017-09-06 X9.3 flare at location 2.}
    \label{X93_S2}
	\vspace{0.1cm}
\end{figure*}

\begin{figure*}
	\centering
	\includegraphics[width=0.8\textwidth]{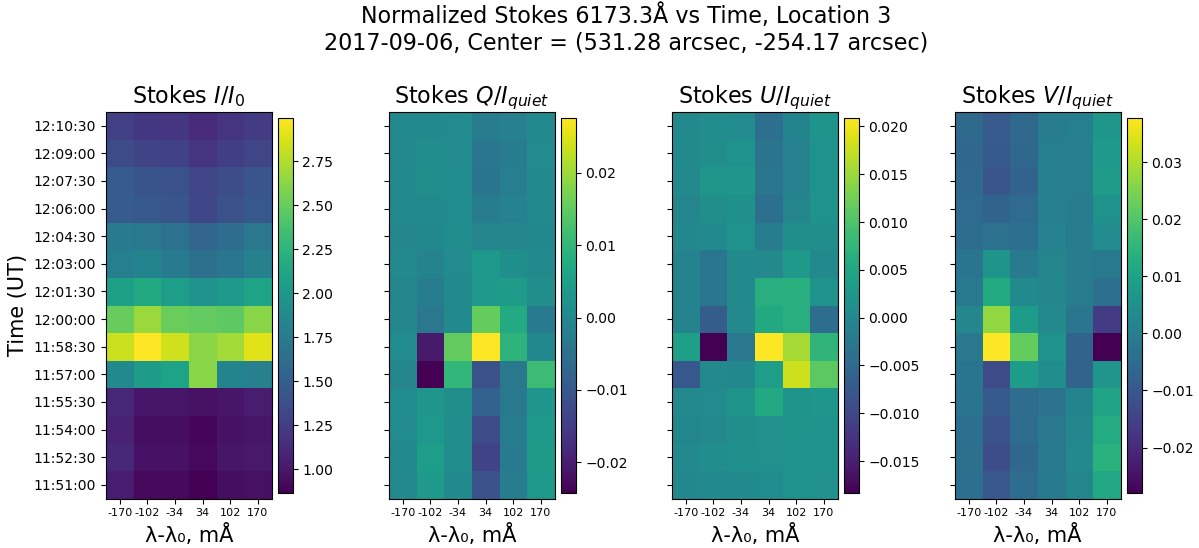}
	\caption{Fe\,I 6173\,\AA~Stokes parameters vs time for the 2017-09-06 X9.3 flare at location 3.}
    \label{X93_S3}
	\vspace{0.1cm}
\end{figure*}

\begin{figure*}
	\centering
	\includegraphics[width=0.8\textwidth]{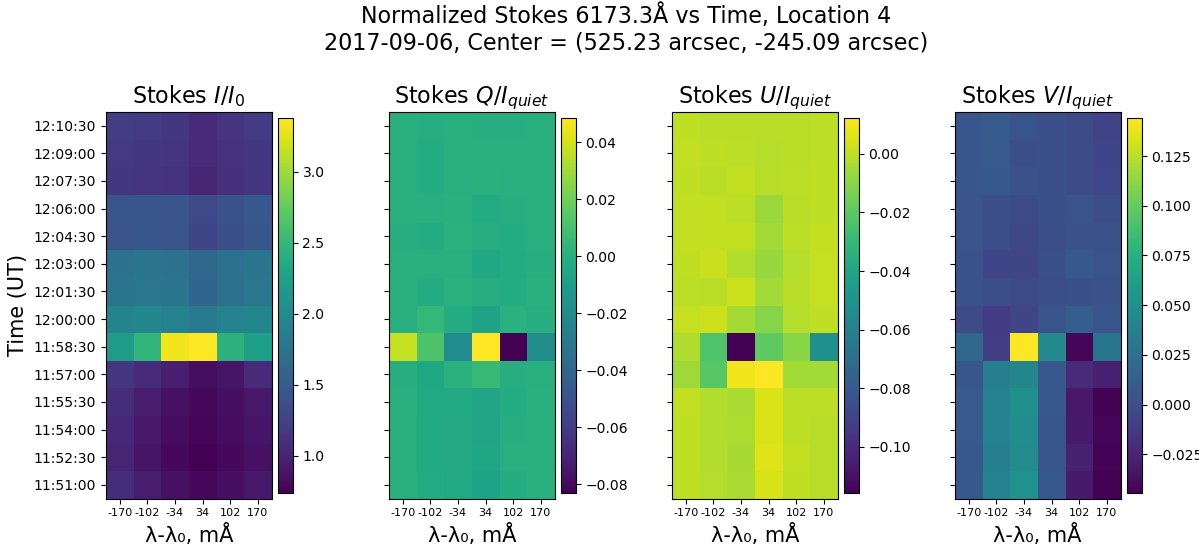}
	\caption{Fe\,I 6173\,\AA~Stokes parameters vs time for the 2017-09-06 X9.3 flare at location 4.}
    \label{X93_S4}
	\vspace{0.1cm}
\end{figure*}

\begin{figure*}
	\centering
	\includegraphics[width=0.8\textwidth]{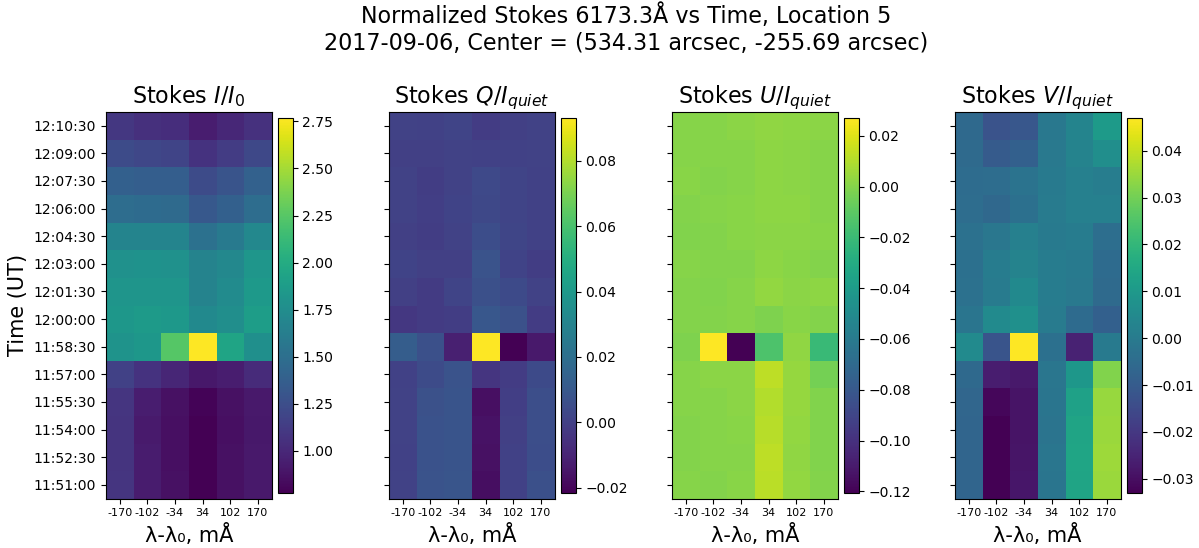}
	\caption{Fe\,I 6173\,\AA~Stokes parameters vs time for the 2017-09-06 X9.3 flare at location 5.}
    \label{X93_S5}
	\vspace{0.1cm}
\end{figure*}

\begin{figure*}
	\centering
	\includegraphics[width=1\textwidth]{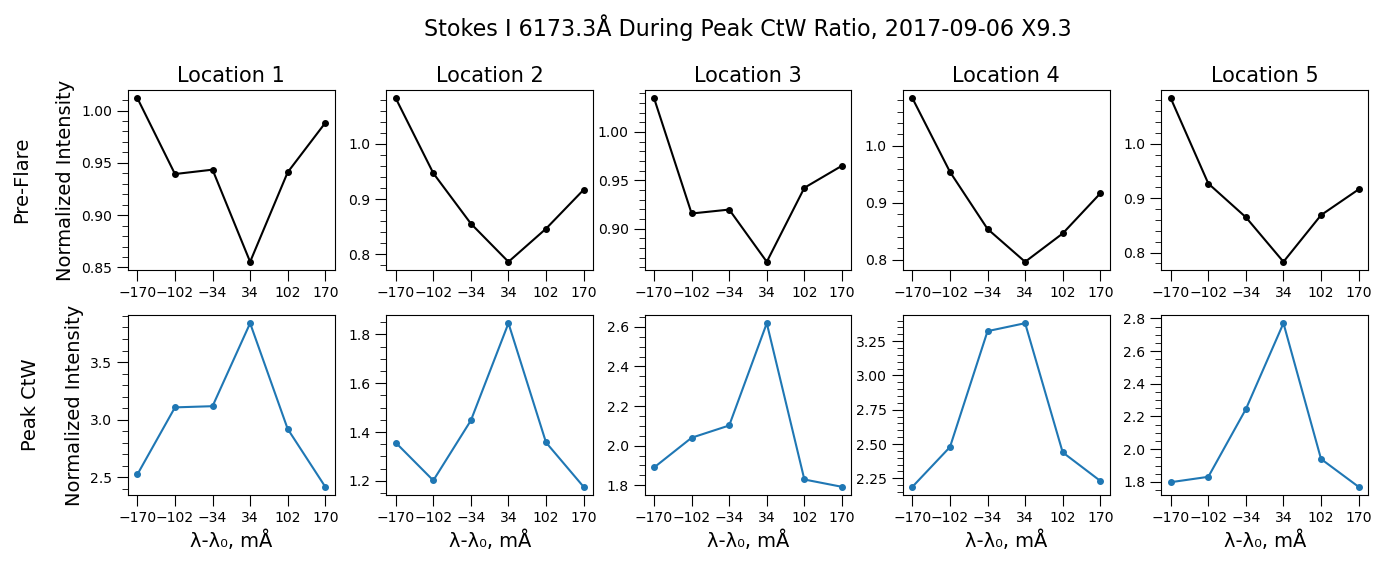}
	\caption{Fe\,I 6173\,\AA~Stokes I during pre-flare (top row) and during maximum CtW (bottom row) for locations 1\,--\,5 for the 2017-09-06 X9.3 flare. Intensity is normalized relative to the mean wing intensity (filtergram channels 1 and 6)}
    \label{X93_SM}
	\vspace{0.1cm}
\end{figure*}

\begin{figure*}
	\centering
	\includegraphics[width=0.55\textwidth]{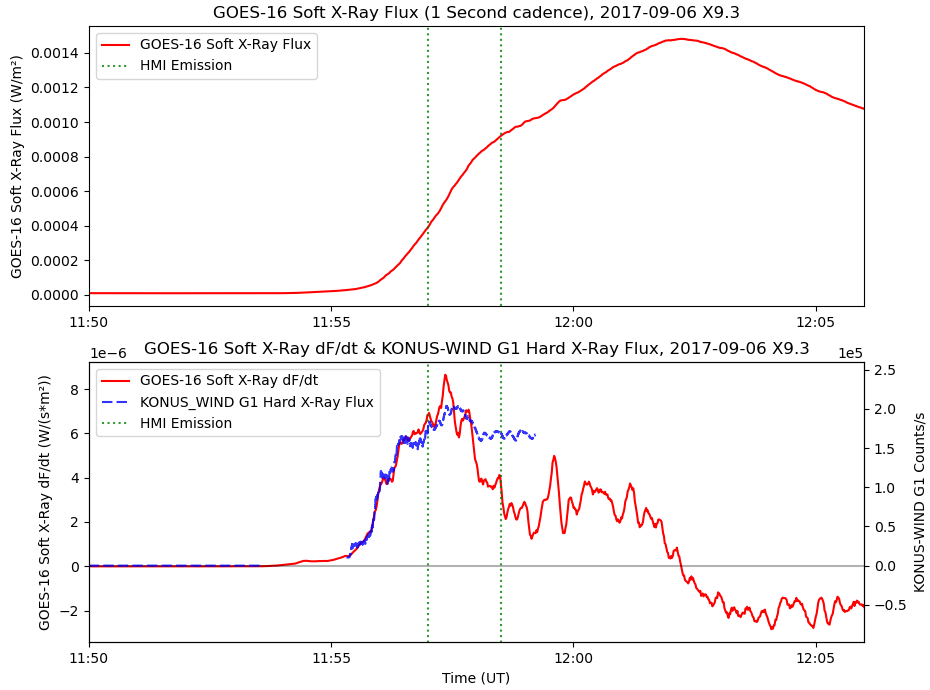}
	\caption{GOES-16 Long Soft X-ray Flux (1.55\,--\,12.4 keV, top panel, solid red), the time derivative of GOES-16 Long Soft X-ray Flux (bottom panel, solid red), and KONUS-WIND channel G1 Hard X-ray (17.8 keV\,--\,75.1 keV, bottom panel, dashed blue) during the 2017-09-06 X9.3 flare. Highlighted are times corresponding to Fe\,I 6173\,\AA~line core emission observed by HMI (vertical dotted green). The X-ray curves are smoothed using a Savitzky-Golay filter.}  \label{X93_Xray}
	\vspace{0.1cm}
\end{figure*}

\subsection{SOL2017-09-06T11:53 X9.3}

Out of the flares which were investigated, the 2017-09-06 X9.3 flare exhibited the highest frequency and intensity of observed line-core emission. Visible in the CtW maps in Figure \ref{X93_CtW}, the line-core emission regions sweep across the umbra and umbra/penumbra boundary of the flaring active region. For detailed analysis, five locations where the line goes into full emission were chosen (shown in Figure \ref{X93_CtW}). These locations will henceforth be referred to as locations 1\,--\,5 with their helioprojective coordinates listed in Figures \ref{X93_S1}-\ref{X93_S5}. The maximum CtW ratio at these locations in order are: 1.69, 1.47, 1.41, 1.78, and 1.50. Between 11:57:00\,UT and 11:58:30\,UT, depending on location, the line-wings were observed to increase in brightness by a factor of 2.47, 1.27, 1.84, 2.21, and 1.78 at locations 1\,--\,5 relative to pre-flare values (Figure \ref{X93_SM}). Meanwhile, the line-core was observed to increase in brightness by a factor of 3.87, 2.01, 2.64, 4.06, and 3.04 at locations 1\,--\,5 relative to pre-flare values. For locations 1, 4, and 5, transient line-core brightening is observed in Stokes I for a single frame as the line goes into full emission. This line-core brightening coincides with brightening in the line-wings. However, due to the 90\,s sampling cadence, these emission events can not be well resolved temporally. For locations 1 and 4, the line then fades to its approximate pre-flare condition over the next ten to twenty minutes. Meanwhile, location 5 observations show a sustained increase in continuum brightness which lasts for roughly five minutes before beginning to fade, indicating heating was sustained for longer relative to other locations. {It should be noted that location 5 is near the tip of a light bridge, so this "sustained brightening" may be caused by the motion of this feature relative to the coordinate system.} Location 4 Stokes I observations also show the line {appears to be} blue-shifted by -33\,m\AA~(-1.6\,km\,s\textsuperscript{-1}) based on Gaussian fitting of the line profile during peak emission relative to both its pre-flare and post-flare state. {However, it is difficult to determine whether this is a true Doppler shift, or an artifact which results from the HMI sampling sequence combined with a short-lived brightening.} In contrast, previous election-beam-based RADYN flare models have only predicted blue-shift velocities up to -0.4\,km\,s\textsuperscript{-1} \citep{Monson2021,Sadykov2020}. 

For locations 2 and 3, core emission {appears to} reach its peak one frame before the line-wings, potentially indicating that the impact affected the higher layers of the photosphere before the lower layers. Similar to locations 1 and 4, the line fades to its pre-flare condition over the next ten to twenty minutes. Stokes Q, U, and V observations for all locations indicate rapid changes in the magnetic field at the time of maximum emission. While in some instances the polarization approaches pre-flare values, there is clear evidence of permanent changes to the magnetic field after the flare, particularly for locations 2, 4, and 5. Based on the KONUS-WIND hard X-ray data and the time derivative of GOES soft-X-ray data shown in Figure \ref{X93_Xray}, these line-core emissions occurred during peak hard X-ray emission. 


\subsection{SOL2017-09-06T08:57 X2.2}
\begin{figure*}[h]
	\centering
	\includegraphics[width=0.8\textwidth]{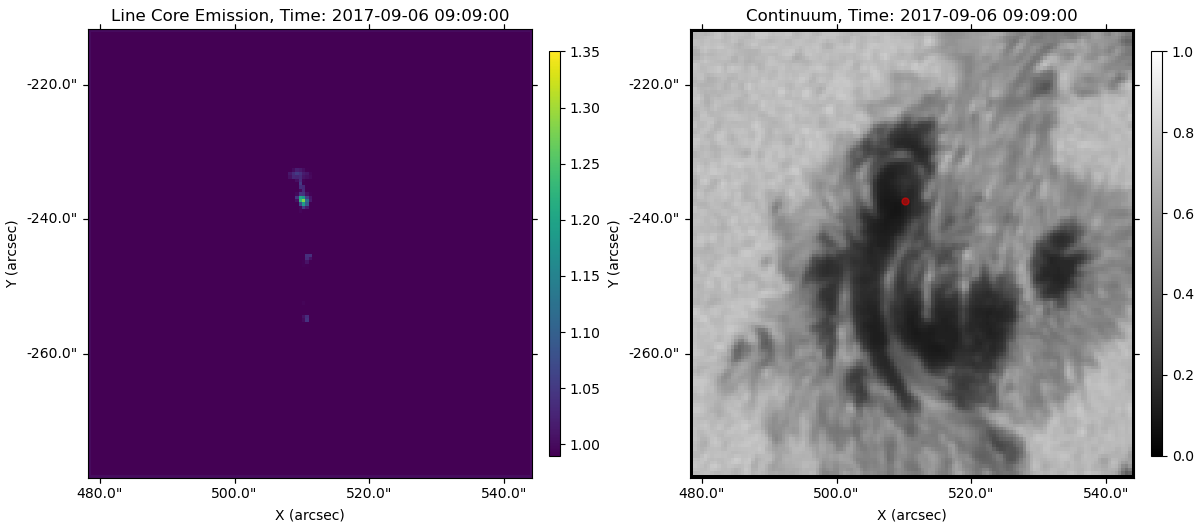}
	\caption{The Fe\,I 6173\,\AA~Stokes I line CtW ratio observed by the HMI instrument during the 2017-09-06 X2.2 flare. The left panel shows CtW ratio during maximum emission. The right panel shows the corresponding normalized HMI continuum map with the maximum observed CtW location indicated by the red dot.}  \label{X22_CtW}
	\vspace{0.1cm}
\end{figure*}

\begin{figure*}
	\centering
	\includegraphics[width=0.8\textwidth]{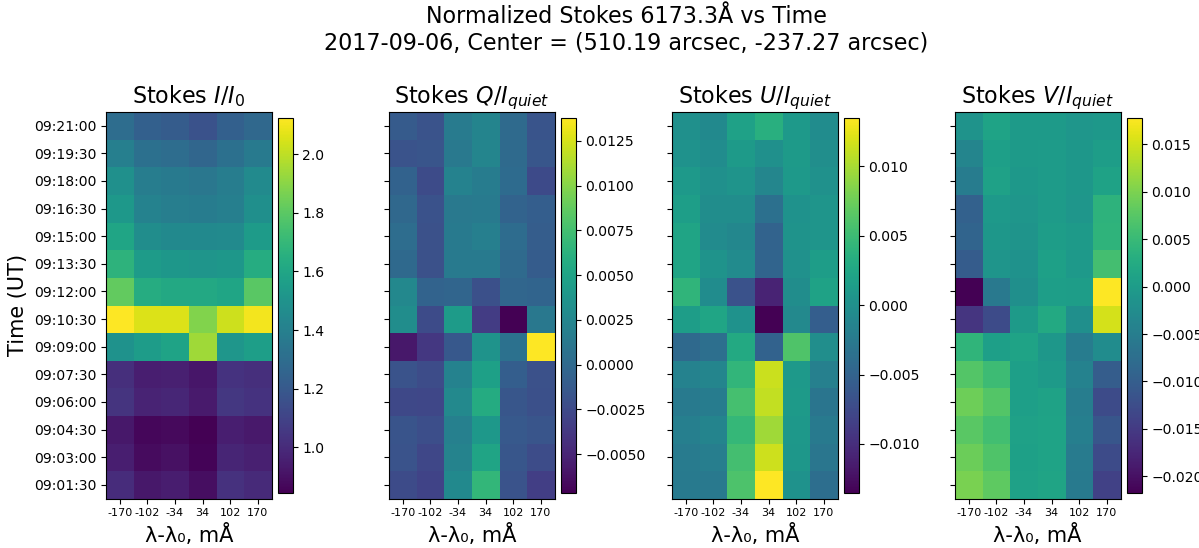}
	\caption{Fe\,I 6173\,\AA~Stokes parameters vs time for the 2017-09-06 X2.2 flare at the location of maximum CtW ratio.}
    \label{X22_S}
	\vspace{0.1cm}
\end{figure*}

\begin{figure*}
	\centering
	\includegraphics[width=0.45\textwidth]{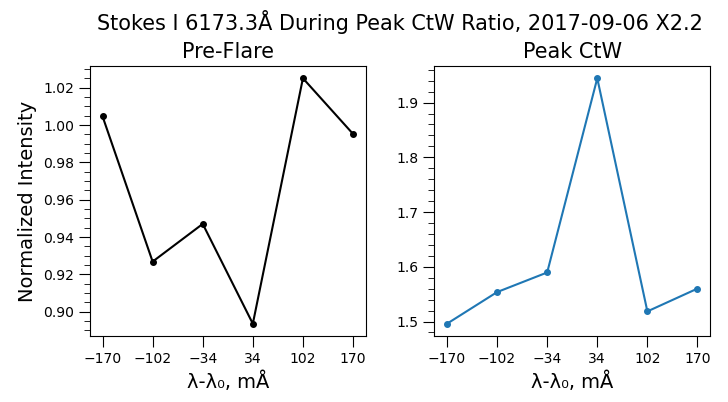}
	\caption{Fe\,I 6173\,\AA~Stokes I during pre-flare (left) and during maximum CtW (right) for the 2017-09-06 X2.2 flare. Intensity is normalized relative to the mean wing intensity (filtergram channels 1 and 6)}
    \label{X22_SM}
	\vspace{0.3cm}
\end{figure*}

\begin{figure*}[h]
	\centering
	\includegraphics[width=0.55\textwidth]{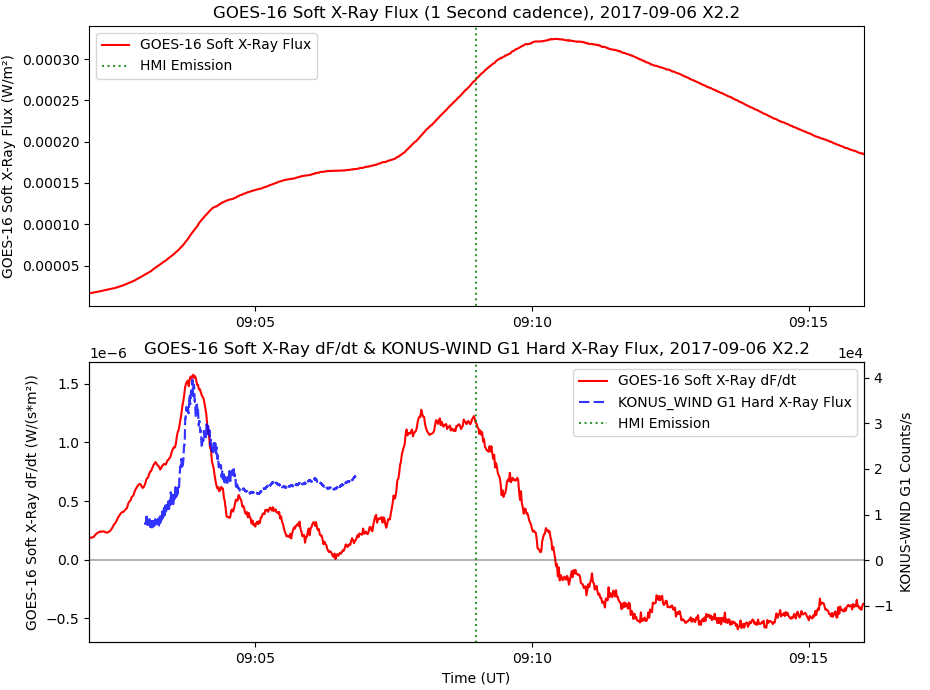}
	\caption{Same as Figure \ref{X93_Xray} for the 2017-09-06 X2.2 flare}  \label{X22_Xray}
	\vspace{0.1cm}
\end{figure*}

The 2017-09-06 X2.2 flare, which took place only three hours before the 2017-09-06 X9.3 flare in the same active region, featured similar umbra-sweeping core-emission (Figure \ref{X22_CtW}). However, a CtW ratio greater than 1.1 was observed at only a single location, at 09:09:00UT, reaching a ratio of 1.29. At this location, the line-core and line-wings are observed to increase in brightness by a factor of 1.92 and 1.53 respectively compared to pre-flare conditions (Figures \ref{X22_S} and \ref{X22_SM}). Line-core emission {appears to} reach its peak one frame before the wings. Additionally, Stokes I decreases at a much more gradual rate when compared with the X9.3 flare data and does not reach its pre-flare state within ten minutes after the time of maximum emission. This indicates that heating lasted significantly longer when compared to the X9.3 flare. Stokes Q, U, and in particular Stokes V are permanently altered after the emission event indicating an inversion in the line of sight (LOS) magnetic field direction along with a decrease in tangential field strength. While line-core emission did not coincide with the first peak in hard X-ray emission observed by KONUS-WIND, it did coincide with the second peak in the soft X-ray time derivative observed by GOES-16 (Figure \ref{X22_Xray}). 

\newpage

\subsection{SOL2022-05-10T13:50 X1.5}

\begin{figure*}[h]
	\centering
	\includegraphics[width=0.8\textwidth]{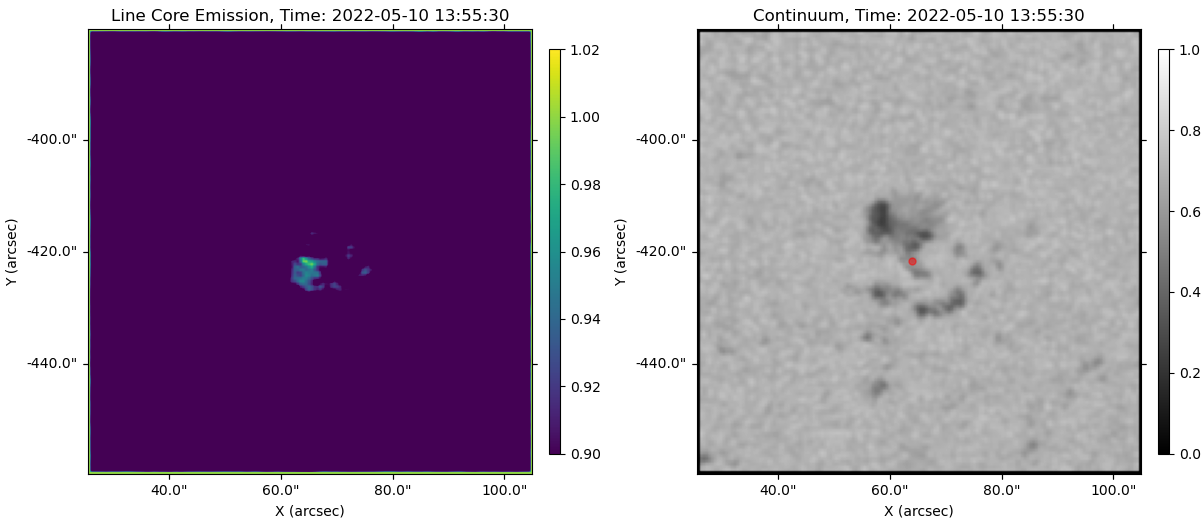}
	\caption{The Fe\,I 6173\,\AA~Stokes I line CtW ratio observed by the HMI instrument during the 2022-05-10 X1.5 flare. The left panel shows CtW ratio during maximum emission. The right panel shows the corresponding normalized HMI continuum map with the maximum observed CtW location indicated by the red dot.}  \label{X15_CtW}
	\vspace{0.1cm}
\end{figure*}

\begin{figure*}[h]
	\centering
	\includegraphics[width=0.8\textwidth]{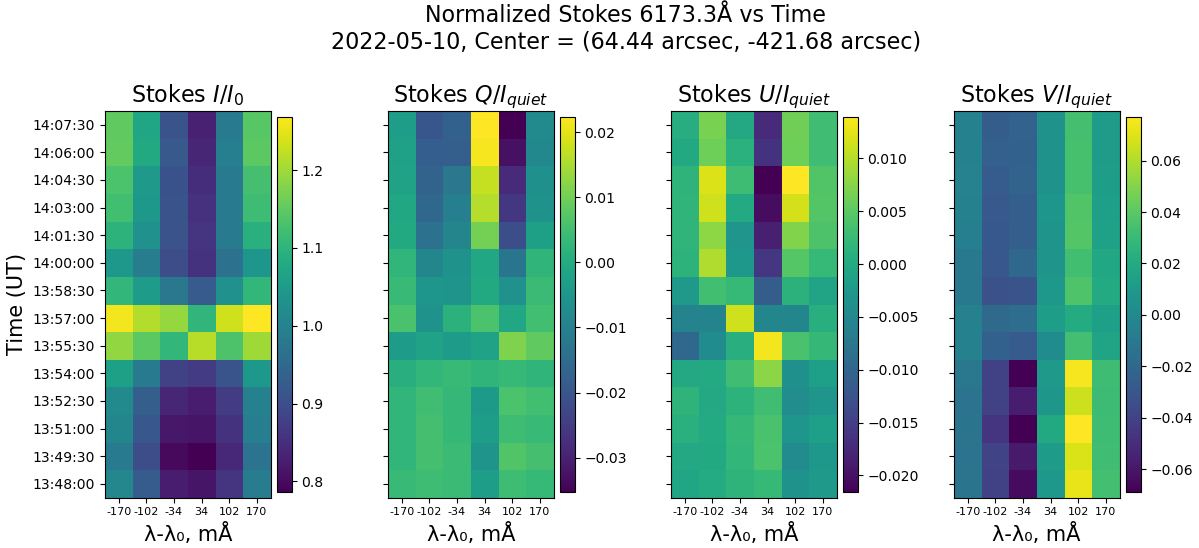}
	\caption{Fe\,I 6173\,\AA~Stokes parameters vs time for the 2022-05-10 X1.5 flare at the location of maximum CtW ratio.}
    \label{X15_S}
	\vspace{0.1cm}
\end{figure*}

\begin{figure*}
	\centering
	\includegraphics[width=0.45\textwidth]{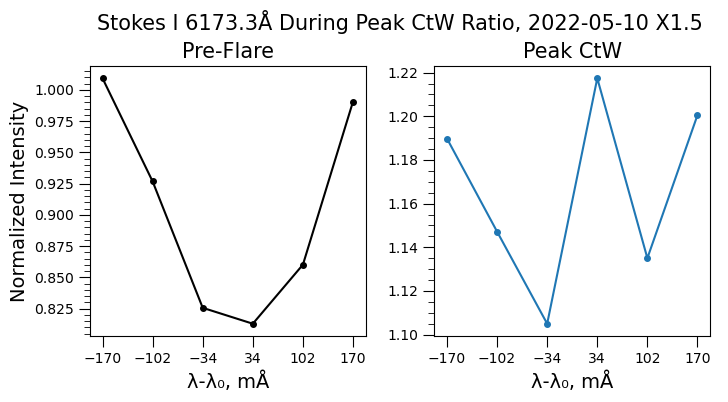}
	\caption{Fe\,I 6173\,\AA~Stokes I during pre-flare (left) and during maximum CtW (right) for the 2022-05-10 X1.5 flare. Intensity is normalized relative to the mean wing intensity (filtergram channels 1 and 6)}
    \label{X15_SM}
	\vspace{0.1cm}
\end{figure*}

\begin{figure*}
	\centering
	\includegraphics[width=0.55\textwidth]{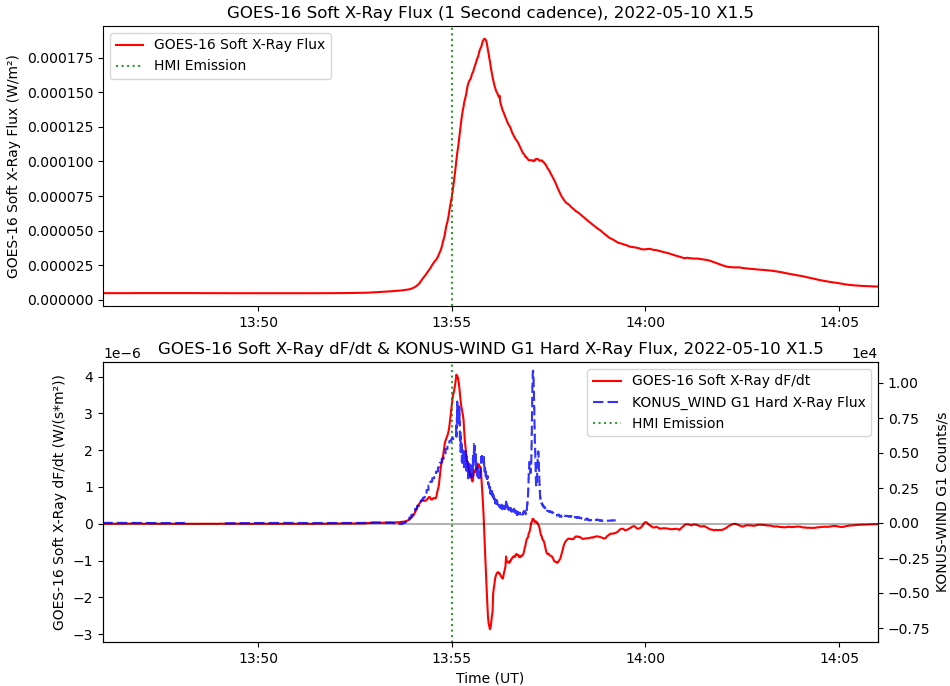}
	\caption{Same as Figure \ref{X93_Xray} for the 2022-05-10 X1.5 flare}  \label{X15_Xray}
	\vspace{0.5cm}
\end{figure*}

Unlike the previously discussed line-core emission events, maximum line-core emission was observed in the penumbra and quiet sun regions during the 2022-05-10 X1.5 flare (Figure \ref{X15_CtW}). Though the maximum CtW ratio of 0.996 is relatively low, it is the greatest ratio observed on the boundary of penumbra and quiet sun out of all investigated events. Maximum line-core emission occurred at 13:55:30 with a 1.42 factor increase in line-core brightness from its pre-flare condition (Figures \ref{X15_S} and \ref{X15_SM}). The line-wings saw a 1.20 factor increase in brightness from pre-flare conditions, and {appears to} reach its peak one frame after peak line-core emission. The line initially fades rapidly over the span of 90\,--\,180\,s, after which the line-wings increase in brightness and stabilize to a value roughly 15\% higher compared to pre-flare conditions. Stokes Q, U, and V are all permanently altered after the emission event indicating an increase in tangential magnetic field strength and a decrease in LOS field strength (Figure \ref{X15_S}). Line-core emission appears to coincide with the first peak in hard X-ray emission observed by KONUS-WIND and the soft X-ray time derivative observed by GOES-16 (Figure \ref{X15_Xray}). 

\newpage

\subsection{SOL2023-12-31T21:36 X5.0}
\begin{figure*}[b]
	\centering
	\includegraphics[width=0.8\textwidth]{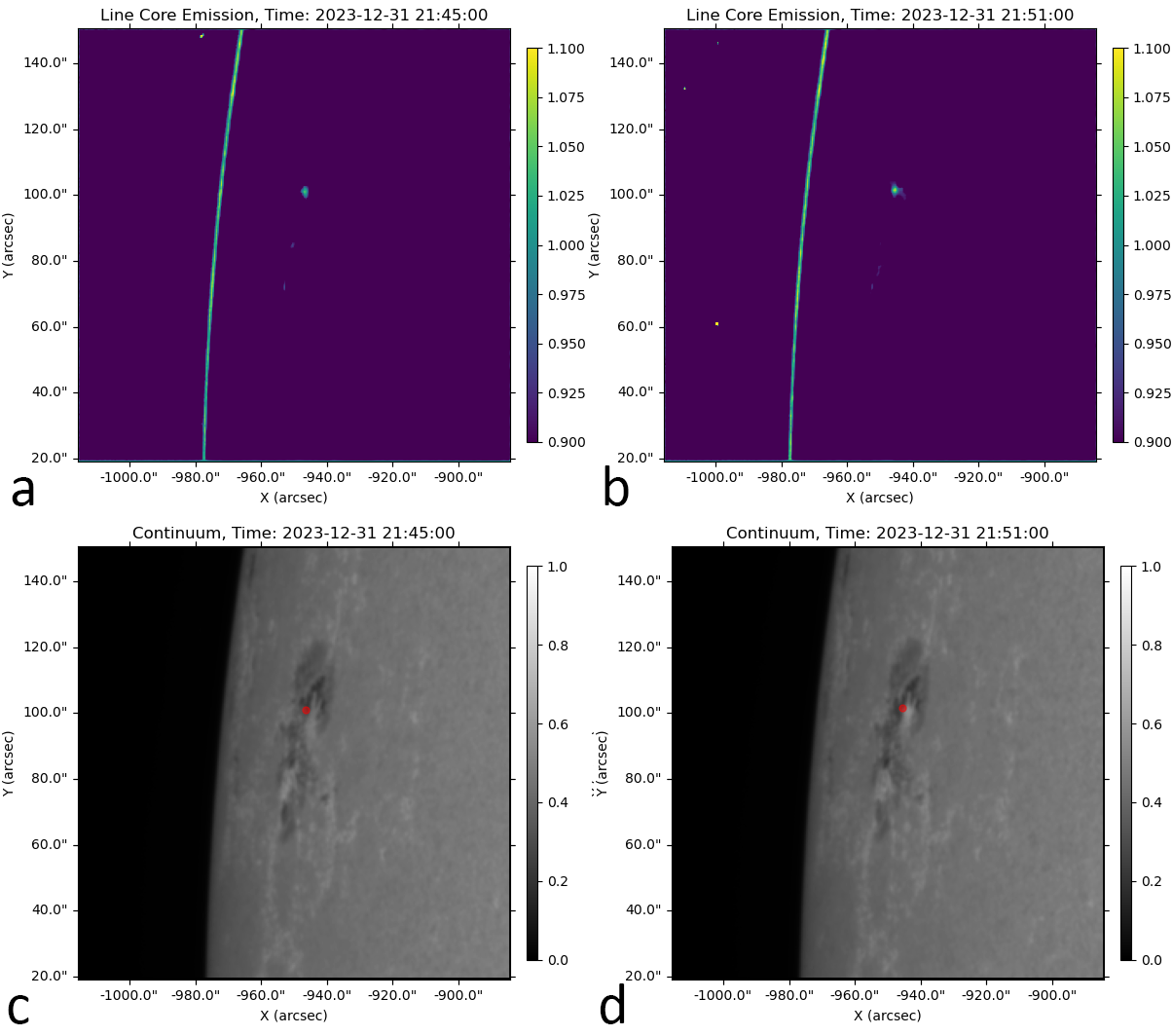}
	\caption{The Fe\,I 6173\,\AA~Stokes I line CtW ratio observed by the HMI instrument during the 2023-12-31 X5.0 flare. Panels a and b show CtW ratio during maximum emission. Panels c and d show the locations of maximum CtW ratio in red on the corresponding normalized HMI continuum maps.}  \label{X50_CtW}
	\vspace{-0.5cm}
\end{figure*}

\begin{figure*}
	\centering
	\includegraphics[width=0.8\textwidth]{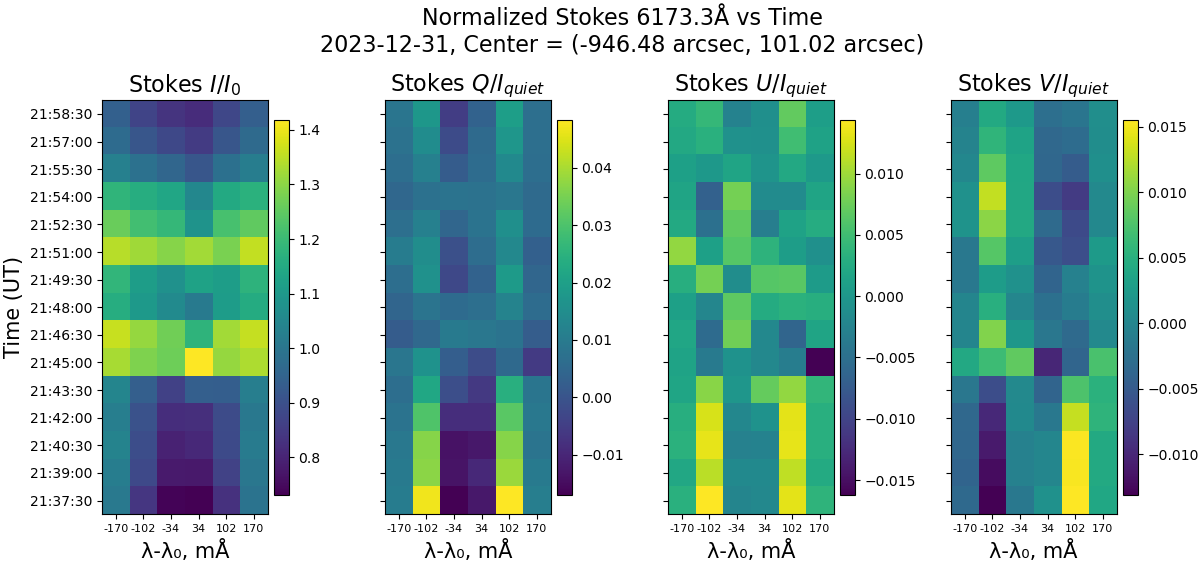}
	\caption{Fe\,I 6173\,\AA~Stokes parameters vs time for the 2023-12-31 X5.0 flare at the location of maximum CtW ratio at t=21:45:00.}
    \label{X50_S1}
	\vspace{-0.35cm}
\end{figure*}

\begin{figure*}
	\centering
	\includegraphics[width=0.8\textwidth]{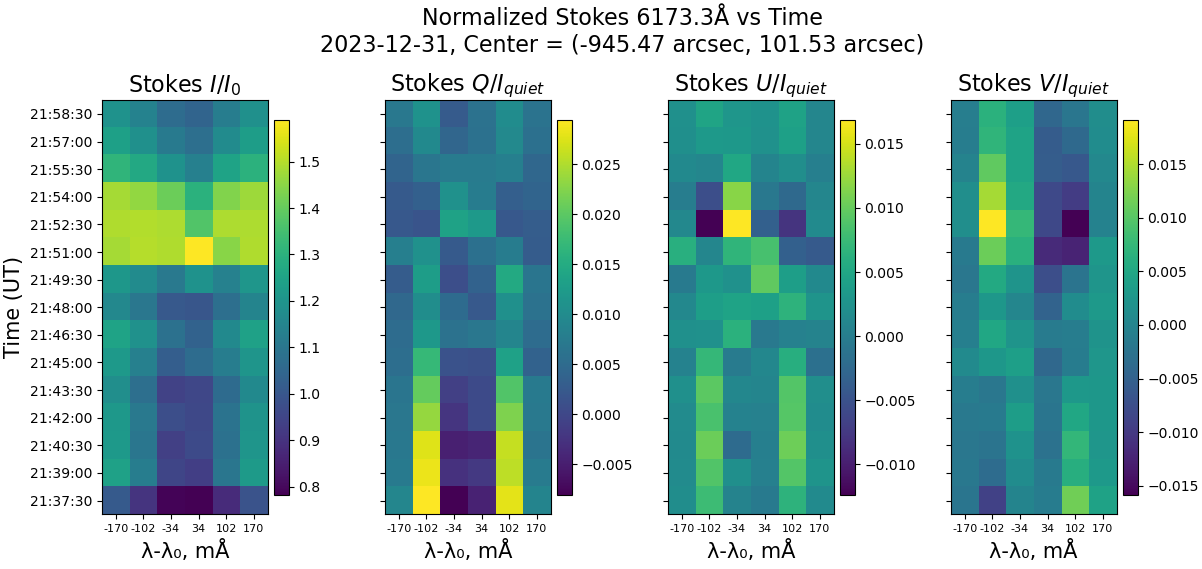}
	\caption{Same as Figure \ref{X50_S1} at t=21:51:00.}
    \label{X50_S2}
	\vspace{0.1cm}
\end{figure*}

\begin{figure*}
	\centering
	\includegraphics[width=0.45\textwidth]{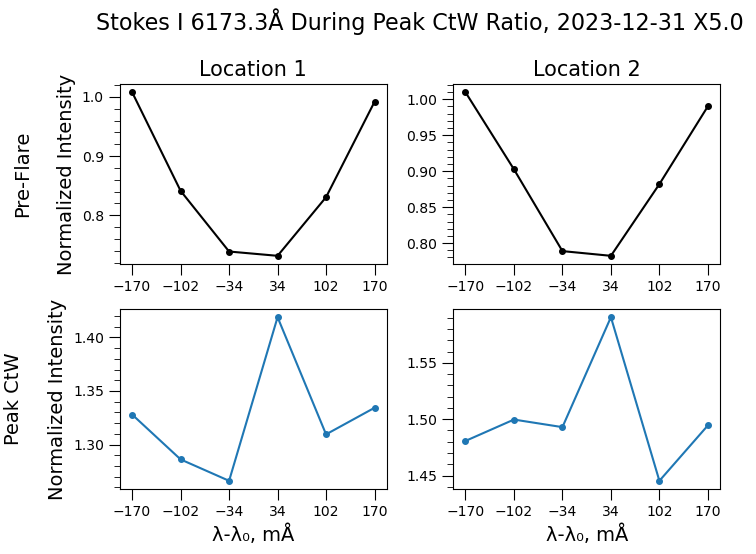}
	\caption{Fe\,I 6173\,\AA~Stokes I during pre-flare (top row) and during maximum CtW (bottom row) for locations 1 and 2 for the 2023-12-31 X5.0 flare. Intensity is normalized relative to the mean wing intensity (filtergram channels 1 and 6)}
    \label{X50_SM}
	\vspace{0.1cm}
\end{figure*}

\begin{figure*}
	\centering
	\includegraphics[width=0.55\textwidth]{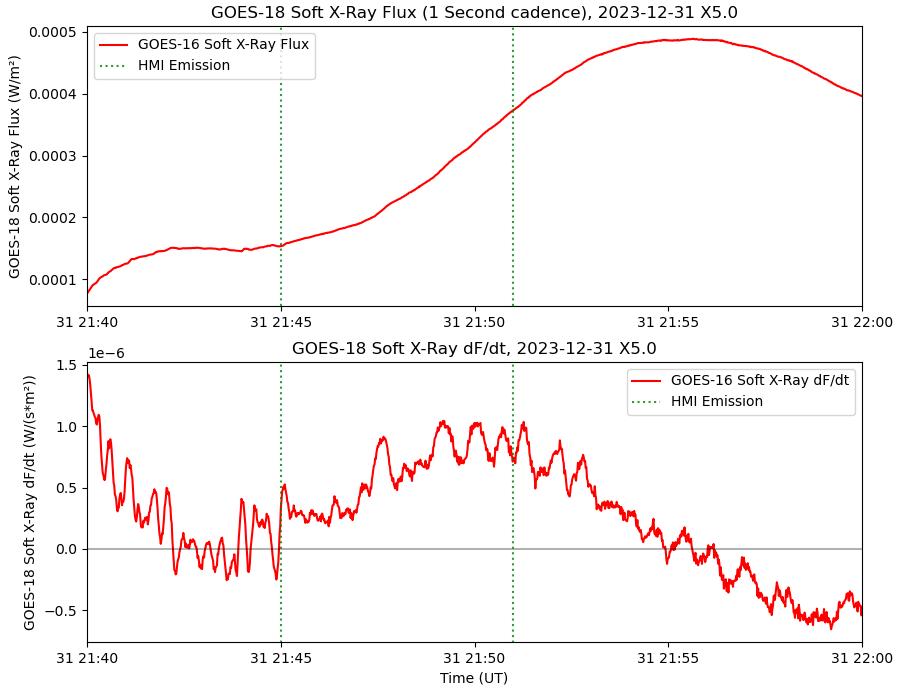}
	\caption{Same as Figure \ref{X93_Xray} for the 2023-12-31 X5.0 flare}  \label{X50_Xray}
	\vspace{-0.3cm}
\end{figure*}

During the 2023-12-31 X5.0 flare, two line-core emission events were observed within an umbral region of the flaring active region at 21:45:00 and 21:51:00 with CtW ratios of 1.05 and 1.08 respectively (Figure \ref{X50_CtW}). Unlike other investigated line-core emission events, observations were taken close to the solar limb, so perspective and optical depth effects likely influence the data. The initial 1.83 factor increase in line-core emission at 21:45:00 occurs for a single frame, and coincides with a 1.33 factor increase in line-wing brightness which lasts for two frames before quickly fading over the next two frames (Figures \ref{X50_S1} and \ref{X50_SM}). At 21:51:00, a second 1.96 factor increase in line-core emission is observed 1.15 arcsec north east from the location of the 21:45:00 line-core emission. This line-core emission also lasts for a single frame and coincides with a 1.49 factor increase in line-wing brightness which lasts for 3 frames before fading (Figures \ref{X50_S2} and \ref{X50_SM}). An increase in brightness of the entire line is also observed at this time in the first line-emission location along with partial line core emission (Figure \ref{X50_S1}). For both locations, large {apparent} red-shifts were also measured: 34\,m\AA~or +1.7\,km\,s\textsuperscript{-1} for the first location and 32\,m\AA~or +1.6\,km\,s\textsuperscript{-1} for the second location. {If these are true Doppler shifts}, the corresponding motions are nearly tangential to the photosphere as these observations were taken very close to the limb.  
Stokes Q, U, and Stokes V are permanently altered after the first emission event indicating a significant weakening in both LOS and tangential magnetic field strength with the direction of the LOS field being inverted. During the second emission event, Stokes Q and V are temporarily strengthened before returning towards their values before the second emission event. KONUS-WIND hard X-ray data is not available for this flare, and though the second line-core emission appears to coincide with the peak in the soft X-ray time derivative observed by GOES-18, the first emission event occurs well before the peak estimated hard X-ray emission (Figure \ref{X50_Xray}).

\newpage

\subsection{SOL2024-02-22T22:08 X6.3}
\begin{figure*}[h]
	\centering
	\includegraphics[width=0.8\textwidth]{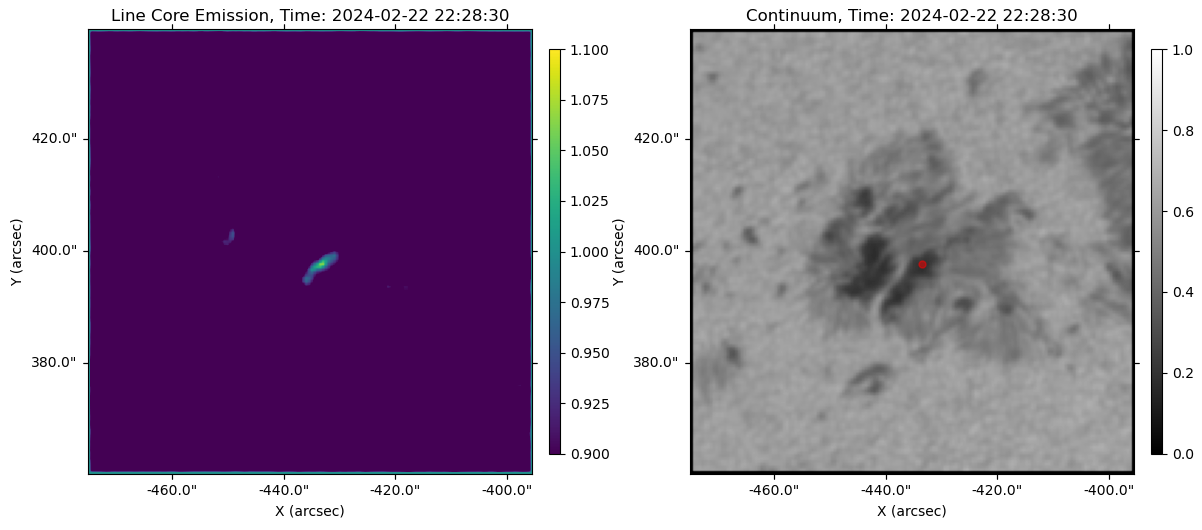}
    \vspace{-0.3cm}
	\caption{The Fe\,I 6173\,\AA~Stokes I line CtW ratio observed by the HMI instrument during the 2024-02-22 X6.3 flare. The left panel shows CtW ratio during maximum emission. The right panel shows the corresponding normalized HMI continuum map with the maximum observed CtW location indicated by the red dot.}  \label{X63_CtW}
	\vspace{-1.8cm}
\end{figure*}

\begin{figure*}[h]
	\centering
	\includegraphics[width=0.8\textwidth]{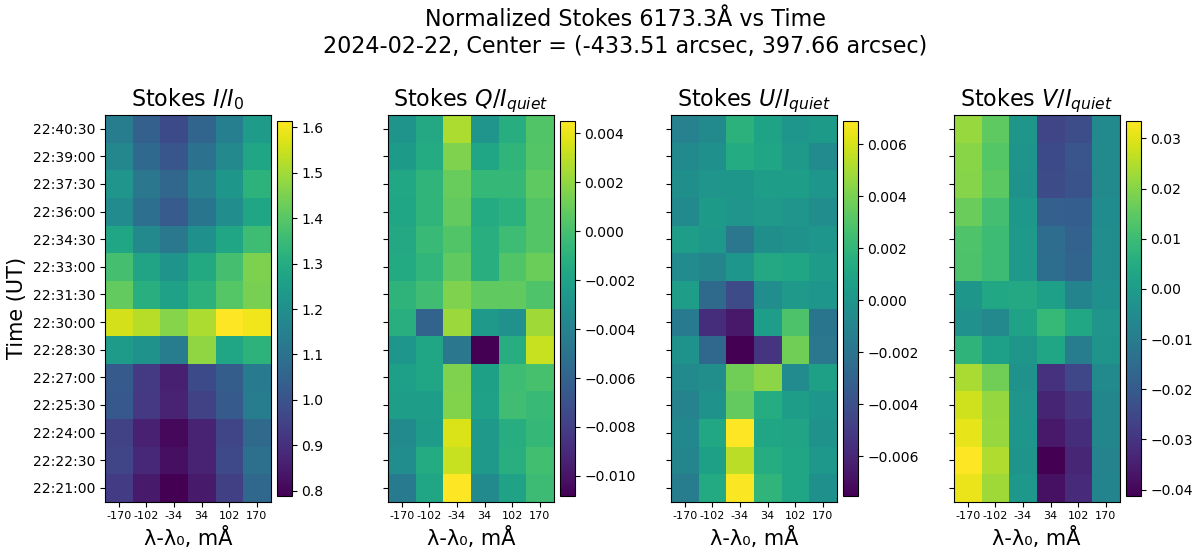}
    \vspace{-0.2cm}
	\caption{Fe\,I 6173\,\AA~Stokes parameters vs time for the 2024-02-22 X6.3 flare at the location of maximum CtW ratio.}
    \label{X63_S}
	\vspace{-0.5cm}
\end{figure*}

\begin{figure*}[h]
	\centering
	\includegraphics[width=0.45\textwidth]{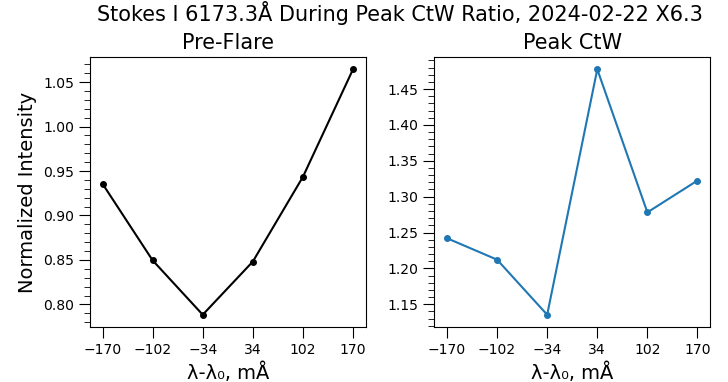}
    \vspace{-0.2cm}
	\caption{Fe\,I 6173\,\AA~Stokes I during pre-flare (left) and during maximum CtW (right) for the 2024-02-22 X6.3 flare. Intensity is normalized relative to the mean wing intensity (filtergram channels 1 and 6)}
    \label{X63_SM}
\end{figure*}

\begin{figure*}
	\centering
	\includegraphics[width=0.55\textwidth]{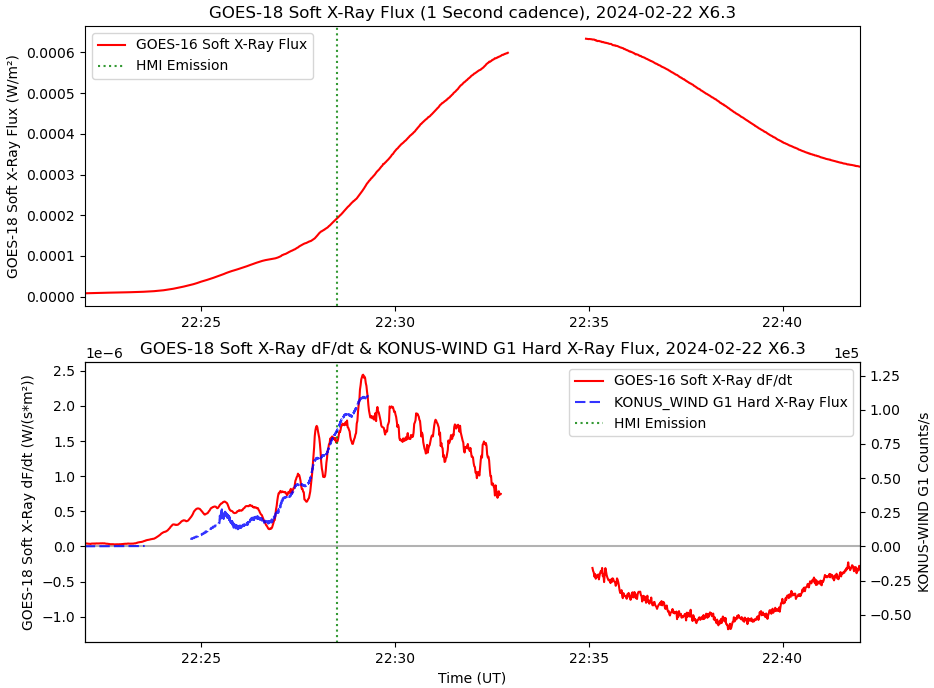}
	\caption{Same as Figure \ref{X93_Xray} for the 2024-02-22 X6.3 flare}  \label{X63_Xray}
\end{figure*}

During the 2024-02-22 X6.3 flare, line-core emission was observed within the umbra of the flaring active region at 22:30:00 with a maximum CtW ratio of 1.06. (Figure \ref{X63_CtW}). At this time, line-core brightness increased by a factor of 1.60 (Figures \ref{X63_S} and \ref{X63_SM}). The line-wings saw a 1.28 factor increase in brightness from pre-flare conditions {and appear to} reach their peak one frame after peak line-core emission. After peak line-wing emission, the line gradually fades and does not reach pre-flare conditions within ten minutes, indicating heating lasted for longer than what was observed for most other investigated events. Stokes Q and V appear to be altered only temporarily with Stokes U demonstrating a more permanent weakening of the tangential magnetic field (Figure \ref{X63_S}). Among the investigated events, this flare exhibited the most significant {apparent} Doppler shift during peak emission with line emission red-shifted by 70\,m\AA~(+3.4\,km\,s\textsuperscript{-1}) relative to both the pre-flare and post-flare profiles. {Assuming these are true Doppler shifts}, this greatly exceeds the maximum 200 m/s down-flow velocity predicted by the electron-beam driven flows in the upper photosphere studied by \cite{Monson2021}, who used the FCHROMA grid of RADYN models. However, \cite{Sadykov2024RADYN} found that proton beams with E\textsubscript{c} = 50 keV and $\delta$ = 5 can produce down-flows with velocities comparable to these HMI observations (2.78\,km\,s\textsuperscript{-1}) at similar heights. Line-core emission appears to occur just before the peak in hard X-ray emission observed by KONUS-WIND and the soft X-ray time derivative observed by GOES-16 (Figure \ref{X63_Xray}).

\newpage

\section{Semi-Empirical Radiative Transfer Modeling of Colder Atmospheres}

The common denominator behind the observed line profile reversals is that they happen within the sunspot umbra or penumbra. There, the photosphere is cooler in comparison with the typical quiet Sun atmospheres, and the intensity of the continuum near the SDO/HMI Fe\,I\,6173\,\AA~line profile is weaker. The question is to what extent such initial condition of the atmosphere favors the development of the Fe\,I line profile reversal.

While there are many radiative hydrodynamic simulations of the heating of the solar atmosphere by electron or proton beams \citep[e.g.][]{Allred2005,2015ApJ...809..104A,Kerr2023,Carlsson2023}, the stratification of the initial atmospheres is mostly quiescent Sun or plage-like. There have been few studies of the impact of the pre-flare stratification on the resulting emission \citep[e.g.,][]{Hong2018ApJ...857L...2H,2018ApJ...856..178P}. An accurate modeling of the beam impact on the umbra- and penumbra-like atmosphere requires the development of a radiatively stable initial atmosphere with a temperature and mass density structure resembling umbra or penumbra semi-empirical models \citep[for example, VAL-S and VAL-R,][]{Fontenla2006}. The development of such initial atmospheres requires a significant effort (due in part to having to account for density depletion and molecular opacities) and is out of the scope of the current paper. Instead, in this work we perform some experiments where we introduce ad-hoc temperature modifications for the existing proton beam simulation solutions \citep{Kerr2023,Sadykov2024RADYN}. While not representing self-consistent temperature evolutions (e.g. radiative heating and cooling, mass motions, height-dependent particle beam energy losses are not considered) in our ad-hoc atmospheres), we do this for illustration to gain a qualitative understanding of how the Fe\,I line profile changes with the change of the initial temperature profile, $T(h)$. Here is our strategy to do this:
\begin{itemize}
    \item Consider a proton beam heating model produced with the radiative hydrodynamics code RADYN \citep{2015ApJ...809..104A,2020ApJ...902...16A,Kerr2023} and analyzed previously by \citet{Sadykov2024RADYN} where the Fe\,I\,6173\,\AA~line profile core demonstrated the strongest emission signatures, yet not leading to a reversal. This model has spectral energy index of $\delta$ = 3, a high low cutoff energy of E\textsubscript{c} = 3\,MeV, and a total energy flux density of {$1\cdot10^{11}$ erg\,s$^{-1}$\,cm$^{-2}$}. The smallest ratio of the line profile depth (defined as the difference between the intensity of the continuum near the line and the smallest intensity along the line) to the continuum intensity near the line was approaching $\sim$0.14 (the corresponding Core-to-Wing ratio of $\approx$ 0.93, described below) for this model \citep[see Table\,1 in][]{Sadykov2024RADYN}, almost leading to the line profile reversal;
    \item Modify the temperature structure $T(h)$ of the initial atmosphere (at t=0\,s) to match the temperature profile $T(h)$ of the VAL-S or VAL-R model within $z = [-50\,-\,420]$\,km and perform a linear transition to the modeled temperature at $z = [420\,-\,1500]$\,km. Record the corresponding $\Delta{}T(h)$ change as a function of height;
    \item Adjust all the following time steps of the model when the flare begins by the same $\Delta{}T(h)$ without changing the other atmospheric parameters (such as densities or bulk gas velocity). Note that the changes will apply only to the -50\,--\,1500\,km portion of the atmosphere;
    \item Compute the Fe\,I\,6173\,\AA~line profiles for each time step by assuming statistical equilibrium conditions for $H$ and $Fe$, a single iteration for calculation of free electrons, and all other species considered in LTE. The broadening of the spectral lines by microturbulence of 2\,km\,s\textsuperscript{-1} is added as well \citep[same as in][]{Sadykov2024RADYN} The calculation of the atomic populations and radiative transfer in statistical equilibrium is done using the RH1.5D radiative transfer code \citep{rybicki1991RHI,rybicki1992RHII,uitenbroek2001RHPRD,pereira2015RH15D}.
\end{itemize}

\begin{figure}[b]
	\centering
	\includegraphics[width=0.95\textwidth]{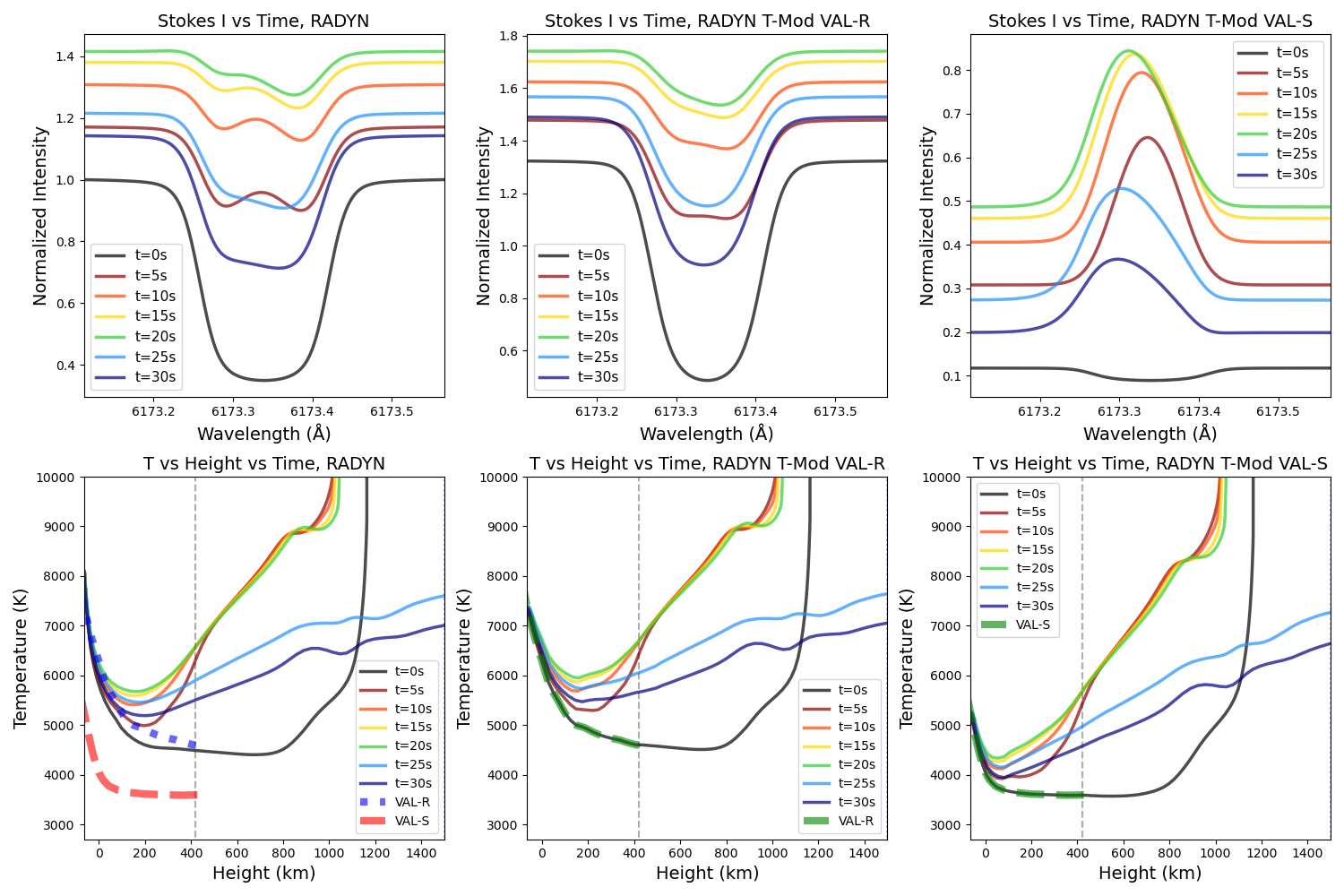}
	\caption{Synthetic Fe\,I 6173\,\AA~Stokes I vs time (top) and temperature vs height vs time (bottom). The left column shows results from the base RADYN proton beam simulation with an energy flux of 10\textsuperscript{11}\,erg\,cm\textsuperscript{-2}\,s\textsuperscript{-1}, a spectral energy index of $\delta$ = 3, a low cutoff energy of E\textsubscript{c} = 3\,MeV, and 20\,s proton beam heating. In addition, the VAL-R (dotted, blue) and VAL-S (dashed, red) models are shown. The middle column shows the result of modifying the base model to match the VAL-R temperature profile (dashed, green), and the right column shows the same as the middle column for the VAL-S model (dashed, green). In the temperature vs height vs time plots, all values to the left of the vertical gray line have been modified such that the t=0 profile matches the VAL-R and VAL-S models, and values to the right are a linear transition from the VAL models to the base RADYN simulation. Stokes I normalization is relative to continuum intensity at t=0 (black) in the base model results.}  \label{RadynVAL}
\end{figure}

\begin{figure}
	\centering
	\includegraphics[width=0.8\textwidth]{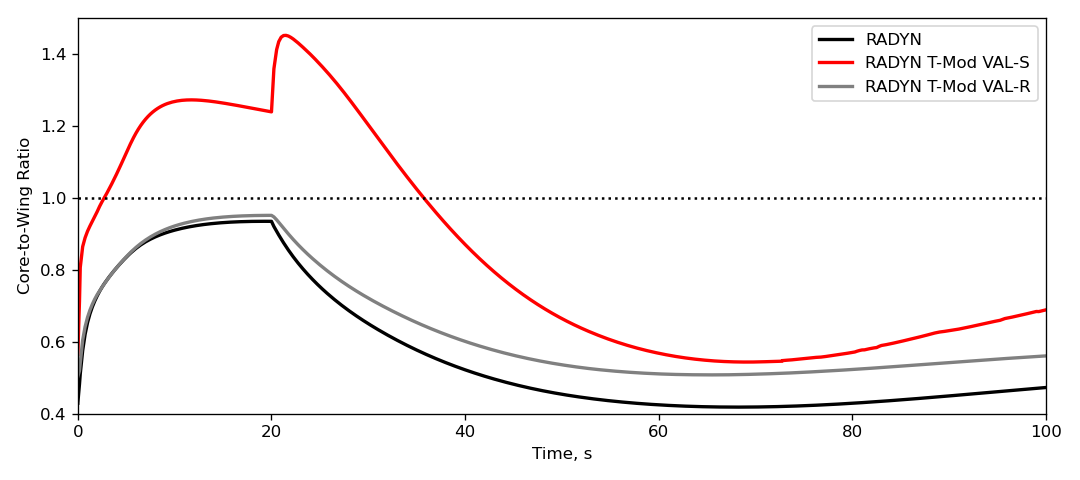}
	\caption{Evolution of the Core-to-Wing Ratios of the Fe\,I 6173\,\AA~spectral line for the considered RADYN model (black), the RADYN model solutions with the ad-hoc temperatures modified to match VAL-S atmosphere (red, RADYN T-Mod VAL-S), and the RADYN model solutions with the temperatures modified to match VAL-R atmosphere (gray, RADYN T-Mod VAL-R). The black dashed line marks the Core-to-Wing ratio of 1.0.}  \label{ctw_models}
\end{figure}

\begin{figure}
	\centering
	\includegraphics[width=1\textwidth]{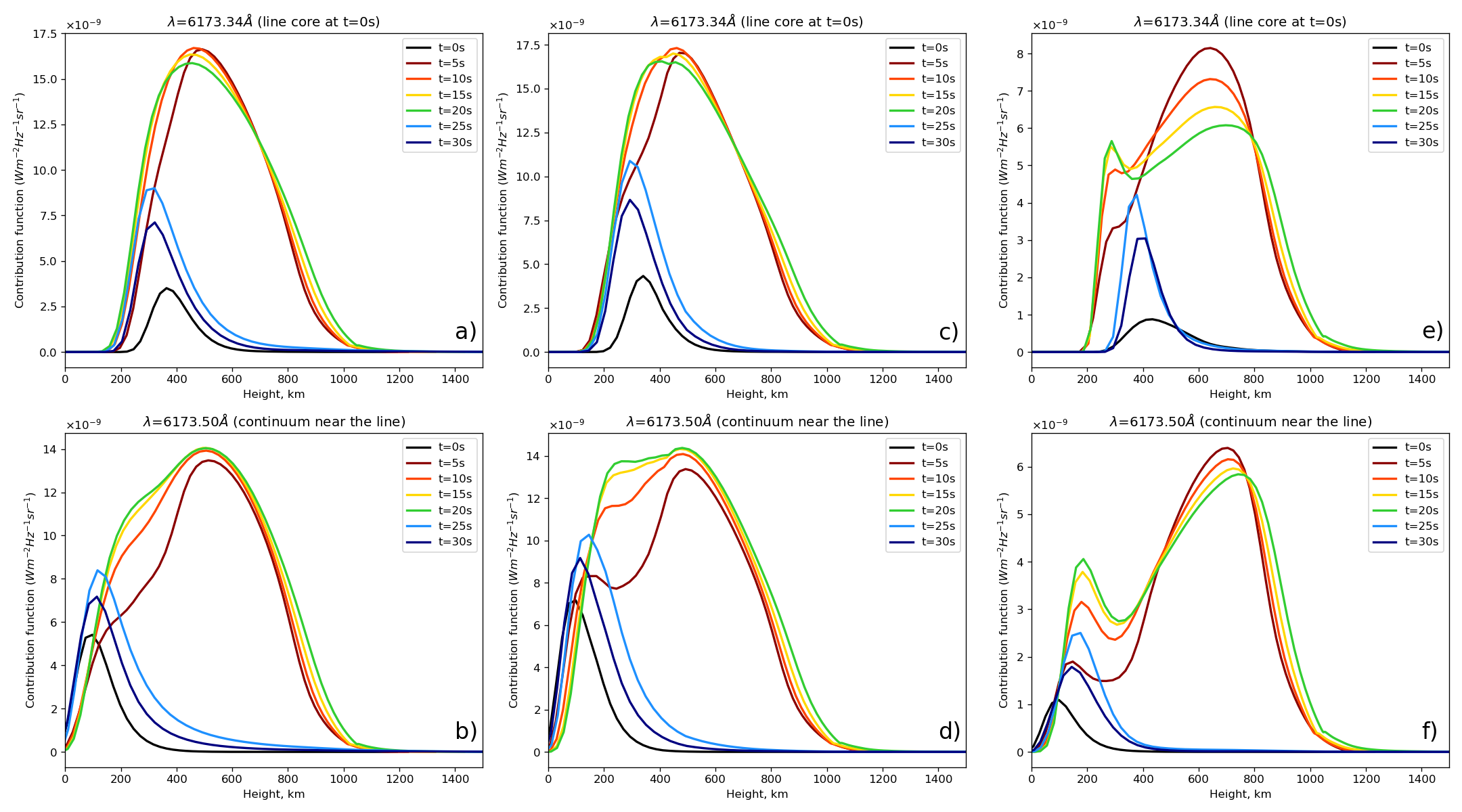}
	\caption{Contribution functions for the Fe\,I\,6173\,\AA~spectral line profiles in the original RADYN simulation atmospheres (panels a and b), RADYN T-Mod VAL-R atmospheres (panels c and d), and RADYN T-Mod VAL-S atmospheres (panels e and f). See Figure~\ref{RadynVAL} for more details. The contribution functions are presented for the time moments similar as in Figure~\ref{RadynVAL} for the rest wavelength of the Fe\,I line ($\lambda$=6173.34\,\AA, left panels) and the wavelength representing the continuum near the line ($\lambda$=6173.50\,\AA, right panels).}  \label{fig:Radyncontrib}
	\vspace{0.1cm}
\end{figure}

While this approach is indeed not self-consistent, it provides a possibility to qualitatively investigate how Fe\,I\,6173\,\AA~line react to the presence of the colder atmospheric temperatures while keeping all other parameters fixed. Thus, we can determine if the emission features result from atmospheres with deep heating and a smaller background radiation field. The results are presented in Figure~\ref{RadynVAL} which demonstrates the line profile at different snapshots of the original model (left column) and of the model where the temperature was adjusted according to VAL-R (hereafter noted as RADYN T-Mod VAL-R; middle column) and VAL-S (hereafter noted as RADYN T-Mod VAL-S; right column) atmospheres. Figure~\ref{ctw_models} demonstrates the evolution of the Core-to-Wing Ratios computed for the simulated line profiles. For the computation of these ratios, the lines are sampled at every snapshot at the wavelengths corresponding to $\pm$34~m\AA\ (core) and  $\pm$170\,m\AA~(wing) assuming the spectral transmission profile of FWHM$\,\approx$\,76\,m\AA\ (effectively corresponding to SDO/HMI). In addition, Figure~\ref{fig:Radyncontrib} demonstrates the contribution functions for the Fe\,I line at $\lambda{}$\,=\,6173.34\,\AA~(the rest wavelength of the Fe\,I line in the unperturbed initial atmosphere, the top row) and for the continuum near the Fe\,I line at $\lambda{}$=6173.50\,\AA~(bottom row).

Overall, the temperature vs height parameter has the greatest influence on the shape of the Stokes I line profile, with lower temperatures above the photosphere consistently resulting in greater line-core emission, with the adjustments based on VAL-S (RADYN T-Mod VAL-S) resulting in the line going into full emission. Interestingly, although the strongest emission in the line profile core for this model exceeds the continuum almost two times (see the top right panel in Figure~\ref{RadynVAL}), the corresponding Core-to-Wing ratio slightly exceeds $\sim$1.4. Such a ratio is comparable to some examples of the observational Core-to-Wing ratios noted in the previous Section (see, for example, Figure~\ref{X93_SM}).

The model based on VAL-R (RADYN T-Mod VAL-R) features slightly higher temperatures within the range of heights of 0\,--\,400\,km. Panels c and d in Figure~\ref{fig:Radyncontrib} indicate that there is a significant contribution to the Fe\,I line formation from these heights, suggesting the possible reason for the less prominent line-core emission. These results qualitatively suggest that the decreased atmospheric temperature in the lower photosphere could potentially favor the reversal of the Fe\,I line profile. {\cite{Jurvcak2018} (Fig. 3) found that heating in the higher photospheric layers was necessary to produce Fe\,I 6301/6302,\AA\ line-core emission observed by Hinode during the September 6, 2017 X9.3 flare. Notably, similar to our VAL-S model, \cite{Jurvcak2018} reported a photospheric temperature of approximately 4100K at $\tau$ = 0 during the post-flare phase. The results presented here are consistent with these findings and additionally indicate that a cooler initial temperature stratification in the lower photosphere enhances the susceptibility of the Fe\,I 6173\,\AA\ line to enter emission under comparable upper-layer heating.} Whether such reversals are exclusive to sunspot umbras or if they could appear in more quiescent Sun-like atmospheric stratification by proton and electron beams of stronger energy fluxes remains to be seen, and will be a subject of future investigations. As an additional note, Figure~\ref{fig:Radyncontrib} indicates that during the heating of the atmosphere by the considered proton beam the formation of the Fe\,I line and the nearby continuum becomes very non-localized (spanning the range of heights of $\sim$100\,--\,1000\,km), with a very significant contribution from the heights $>500$\,km. Moreover, the nearby continuum features greater contributions from heights $<200$\,km compared to the line-core both before and during heating. More detailed studies of the formation of Fe\,I line profile need to be investigated in the following up, self-consistent, modeling efforts.

\section{Summary}

This study presents a detailed spectro-polarimetric analysis of Fe\,I 6173\,\AA~line emission during five WLFs observed by SDO HMI between 2017 and 2024. For each flare event, a 19.5 to 22.5 minute time-series of 90\,s cadence HMI Stokes data is compiled while accounting for differential rotation effects to allow for temporal analysis. Of the investigated flares, the SOL2017-09-06T11:53 X9.3 flare exhibited the highest frequency and intensity of observed line emission. This emission occurred primarily above sunspot umbra and umbra/penumbra boundaries, with the line core-to-wing ratio reaching as high as 1.78. Meanwhile, the strongest observed line-core emission above penumbral or quiet sun regions out of the investigated events, observed during the 2022-05-10 X1.5 flare, has a relatively small core-to-wing ratio of 0.996. For all instances, line core emission consistently lasts for a single 90\,s frame and appears either concurrent with or delayed by one frame relative to continuum brightening. In cases with delayed continuum brightening, the associated flare photospheric impact affects higher layers of the photosphere first. {However, due to the 90\,s sampling cadence and HMI observation sequence, there is some ambiguity regarding the exact timing of events. Large Doppler shifts of 1\,--\,4 km\,s\textsuperscript{-1} are observed during peak line-core emission for some flares. However, it is difficult to determine the accuracy of these measurements because of the fast variations in the flare white-light cores.}Following the emission event, continuum intensity fades toward its pre-flare state over the next three to twenty minutes depending on the duration of heating. Lasting changes to Stokes Q, U, and or V were observed for all emission events indicating permanent changes in the magnetic field at these locations. These emissions coincided with local maxima in hard X-ray emission observed by Konus-Wind along with local maxima in the time derivative of soft X-ray emission observed by GOES satellites, suggesting the line core emissions may be concurrent with the particle beam precipitation. Again, due to the 90\,s sampling cadence, it is difficult to determine the exact relative timing of the two events.  

The recent RADYN proton beam flare simulations utilized photospheric temperature profiles characteristic of plage-like atmospheres. Given the disparity between proton beam cutoff energies that encourage line-core emission and those that drive sunquakes in these models, the observations suggest that cooler umbral temperature profiles may enable stronger line-core emission at cutoff energies associated with intense sunquakes. To test to what extent cooler photospheric temperatures result in increased line-core emission for a given proton beam cutoff energy, we perform semi-empirical radiative transfer modeling of the Fe\,I 6173\,\AA~line. This is done by modifying an existing plage-like RADYN proton beam-driven flare simulation. Using modified temperature profiles corresponding to the VAL-S umbra model and VAL-R penumbra model at the initial time step of the simulation, the temperature evolution of the RADYN flare was changed whilst keeping other atmospheric properties fixed. While the original quiet sun and penumbra models feature only partial line core emission, the results indicate that umbral photospheric temperature profiles result in full reversal of the Fe\,I 6173\,\AA~line. While these penumbral and umbral models are physically inconsistent, they demonstrate that lowering photospheric temperatures while maintaining other parameters may result in more favorable conditions for Fe\,I 6173\,\AA~line reversal during proton beam impacts. This motivates the construction of penumbral and umbral models to model these features more self-consistently. Those future simulations can be used to confirm the hypothesis that we present in this manuscript. Our study also demonstrates that the underlying pre-flare atmosphere is a key factor when performing model-data comparisons. 

\vspace{0.5cm}

\section*{Acknowledgment}
This work was partially supported by the NSF grant 1916509. VMS acknowledges the NSF FDSS grant 1936361. GSK acknowledges support from a NASA Early Career Investigator Program award (grant \# 80NSSC21K0460). The authors acknowledge the use of the Overleaf Writeful and TeXGPT models for sentence structure and prose suggestions.

\newpage


\end{document}